\documentclass[12pt]{article}

\input{epsf.tex}

\begin{document}

\begin{center}
{\bf REANALYSIS OF THE PROCESS OF THE CASCADE GAMMA DECAY OF $^{198}$Au
COMPOUND STATE}\\\end{center}
\begin{center}
{ A.M. Sukhovoj, V.A. Khitrov}\\
\end{center}\begin{center}
{\it Frank Laboratory of Neutron Physics, Joint Institute
for Nuclear Research, 141980, Dubna, Russia}\\
\end{center}\begin{center}
{ B.E. Crawford, S.L. Stephenson}\\
{\it Gettysburg College, 300 N. Washington Street. Gettysburg PA 17325, USA}
\end{center}
\begin{abstract}
To further study the "pygmy" resonance phenomena in the photon strength 
function, we reanalyzed the two-step cascade data for the target nucleus
$^{197}$Au using the Dubna group approach.
 The range of obtained values allows for meaningful conclusions:
the level density at low excitation energy shows a step-like behavior;
the electric dipole photon strength function has a broad maximum around
$E_{\gamma}=5$ MeV and is not typical of a "pygmy" resonance;
the level density below $B_{n}$ also demonstrates step-like behavior.
\end{abstract}
\section{Introduction}

The main task of an experiment in the low energy ($E_{ex} {_<^{\sim}}10$ MeV)
physics is to study the influence of structure of excited levels of a nucleus
on the parameters measured in the experiment, for example of the process of
cascade gamma decay.
After that -- the extraction of dynamics of intranuclear interactions out of
these experimental values and their theoretical interpretation with the
development of theoretical models required in practice \cite{RIPL} for the
nuclear parameters used in this case.
This is necessary, in particular, in order to obtain the maximum realistic
evaluation of cross sections of interactions of neutrons with nuclei
necessary in practice. This is important especially for actinides where the
existing models of level density can not \cite{Maslov} provide for the maximum
reliable and accurate evaluation of cross sections with the absolute minimum
of accepted hypotheses and assumptions.

In the stated analysis cycle the evident insufficiency of experiments sensitive
to the structure of nucleus in the widest region of its excitations still
remains the main problem.
At present, the co-existence, interaction and defining influence of the nuclear
excitations of qualitatively differing types, namely multiquasi-particle and
vibration ones, on the structure of nuclei give no rise to doubt.
This is the main conclusion of such fundamental models of nucleus as
different variants of IBM and QPNM.
Unfortunately, the majority of experiments carried out so far give direct
and quite reliable information on the structure of nucleus only for too
small energies of its excitation. In practice, for example, this region in
the even-odd heavy nucleus \cite{W185} is still restricted to the interval
of excitation energies of the order of 2 MeV.
Therefore, there are no direct methods to determine level densities, first
of all, at higher excitation energies.
The mentioned parameter of a nucleus, like probability of the gamma quantum
emission in the whole region lower than the neutron binding energy or
nucleon products of a reaction may only be determined from indirect experiments.
Mainly, such analysis uses the spectra of products of a nuclear reaction
measured by single detectors. Their amplitude depends both on the number
of excited levels and on partial widths of the emission of nuclear reaction
products according to the given decay channel of the initial state of an
excited nucleus.

The situation is also complicated by the fact that the measured ordinary
spectra of one-step reactions are mainly determined by the product of level
density of a nucleus on the probability of emission of their products.
As a result of strong correlation of these parameters, the transfer of
inevitable errors in determination of the spectrum intensity to the unknown
values increases abruptly their uncertainties. This circumstance completely
excludes a possibility of simultaneous experimental determination,
for example of reliable values of level density $\rho$ in a fixed interval
of their spins or of radiative strength functions
$f=\Gamma_{\lambda i}/(E_{\gamma}^3\times D_{\lambda})$
of cascade gamma transitions, without attraction of any model notions:
first of all, without the hypothesis \cite{Bohr}, which is a basic one
for the analysis of all the experiments carried out earlier, on the
indepedence of cross section of the inverse reaction on the excitation
energy of the final nucleus.

Potentially, the task of simultaneous determination of $\rho$ and $k$
could be solved for gamma decay of any excited level $\lambda$
of an arbitrary nucleus with the mass $A$ with any mean spacing $D_{\lambda}$
between them at the combining of experimental data of different
experiments, for example of the cross sections of radiative neutron capture
and spectra of gamma rays occurring simultaneously. However, there are no
practical achievements in this direction so far.

A fundamentally new method to solve the problem under consideration was
realized for the first time \cite{PTE} in Dubna. The analysis of intensities
of two-step  cascades of radiative capture of thermal neutrons in the energy
intervals $\Delta E$, fixed according to the method \cite{Prim}, of their
intermediate levels $E_i=B_n-E_1$
\begin{equation}
I_{\gamma\gamma}(E_1)=\sum_{\lambda ,f}\sum_{i}\frac{\Gamma_{\lambda i}}
{\Gamma_{\lambda}}\frac{\Gamma_{if}}{\Gamma_i}=\sum_{\lambda ,f} 
\frac{\Gamma_{\lambda
i}}{<\Gamma_{\lambda i}> m_{\lambda i}} n_{\lambda 
i}\frac{\Gamma_{if}}{<\Gamma_{if}> m_{if}}. 
\end{equation} 
made it possible for the first time to determine $\rho$ and $k$ simultaneously
and without a model: in the initial variant - on assumption of the
independence of partial radiative widths $\Gamma$ on the excitation energy
of the studied nucleus (i.e. using the hypothesis \cite{Bohr}); in the modern
one \cite{PEPAN-2005} - completely without using it. The indubitable advantage
of such experiment is also the circumstence that for any interval of excitation
energies $\Delta E$ the number  $m=\rho \Delta E$ (or $n$)of levels is fixed
by the spin window assigned by an experiment.
At the same time systematic errors $\delta \rho$ and $\delta k$
of the determined parameters are restricted very much by the type of spectra
measured in the experiment (in comparison with other methods of similar
experiments) and have a quite acceptable value \cite{TSC-err} for practically
attainable systematic errors 
$\delta I_{\gamma\gamma}$.

Later, similar experiments were carried out in Riga, \v{R}e\v{z}, Los Alamos,
Budapest \cite{Riga,Sn118,LA172,Bud57} and started in Dalat.
However, the conclusions of different groups on parameters of the cascade
gamma-decay process differ fundamentally depending on the method of
experimental data processing used by the authors. This difference is of
purely technical character and may lead to the appearance of false conclusions
about the process under investigation only if one does not take into account
strong correlation in expression (1)
of the unknown parameters $\rho$ and $k$ in the analysis.
Errors increase particularly in conclusions of the analysis of experiment
when anticorrelations of cascade intensities are neglected with the energies
of their primary and secondary gamma transitions located in one and the same
interval $\Delta E$ of each experimental spectrum.

\section{Main tasks and problems of model-free determination of $\rho$ and $k$}

Currently, the method \cite{PEPAN-2005} is the only source to obtain reliable
data on $\rho$ and $k$ for any compound nucleus if only the experimental
conditions allow one to limit the energy spread of the excited levels $\lambda$
to the interval of the order of 1-3 keV and less.
However, in contrast to other already existing methods, here a single-valued
determination of the unknown parameters is impossible in principle.

For example, for cascade gamma decay at the neutron capture in resonance the
information on experimental intensities of two-step cascades, their density
and number of known low-lying levels is available.
However, it is impossible to obtain unambiguity in determination of $\rho$
and $k$ even using \cite{PEPAN-2005} such additional information as the total
(or only cascade) population of levels in the low half of neutron binding
energy. 

A very essential limitation of the region of possible values of level densities
and radiative strength functions is provided by non-linearity of equation (1)
relative to $\rho$ and $k$. The non-linearity effect occurs only if its half
is extracted \cite{Prim} out of the experimental spectrum. The half equals to
the summarized intensity of two-step cascades, which excite intermediate
levels in each given interval of their energy.
This very operation during the data processing decreases the interval of
values of $\rho$ and $k$, which accurately reproduce $I_{\gamma\gamma}$ from
absolutely non-informative ones \cite{Oslo-err} to practically accurate values
of \cite{TSC-err} $\rho$ and $k$ suitable for comparison with the theory.
That is why the analysis \cite{PEPAN-2005} makes it possible to obtain the
maximum realistic notions on the dynamics of the process of cascade gamma decay
of any nucleus. However, the existing quite serious discrepancies between the
data on $\rho$ and $k$
from the technique \cite{PEPAN-2005} and the technique, which is used for a 
long time, to extract level densities from evaporative spectra point to the
necessity of serious comprehensive analysis of both sources of systematic
errors in the compared experiments and search of factors, which may influence
essentially, in the first place, the determined values of level density.

At present, the problem of studying the influence of structure of wave
functions of levels connected by a cascade on its intensity takes on special
significance in determination
\cite{PEPAN-2005} of $\rho$ and $k$. It is of special importance
for heavy odd-odd compound nuclei where, due to the insufficiency of
experimental data on gamma ray spectra in the region $E_\gamma \sim 0.5B_n$,
we failed to estimate the degree of discrepancy of radiative strength
functions of primary and secondary gamma transitions of one and the same
energy and multipolarity.
It is also true for the nuclei maximally close to the actinide region for
preliminary evaluation of the conditions, which may distinctly distort the
values of $\rho$ and $k$ obtained with the help of method \cite{PEPAN-2005}.

The models \cite{RIPL} of radiative strength functions surpass essentially
with regard to the extent of working over and account of structure of a
nucleus of the level density models.
The second ones take into account in an explicit form the existence of two
fundamentally different types of nuclear excitations, the first ones still
use only excitations of the fermion type \cite{RIPL}. The accumulated data
set for each of the nuclei studied in \cite{Meth1} points to the necessity,
at least, of phenomenological inclusion of the contribution from excitations
(or the corresponding components of wave functions)of vibration type into
radiative strength function models.

Taking into account all these factors, the maximum complete data analysis
for the compound nucleus $^{198}$Au is of primary interest.

\section{Properties of the cascade gamma decay of the $^{198}$Au compound state} 

Experimental determination of the total gamma ray spectrum of the radiative
neutron capture and its interpretation in the framework of the present-day
notions made the authors \cite{Bar198} conclude that in this nucleus the
gamma decay is ``anomalous": at the gamma transition energy of about 5 MeV
the so-called ``pygmy- resonance" \cite{Igash} manifests itself in the radiative
strength function. This interpretation of the observed increase of the
strength function of emission of the corresponding gamma quanta still
remains and it became a subject of investigation in \cite{Pra198}. 
Notions on the ``anomality" of gamma decay have been obtained and remain only
in the framework of the condition that the level density has been determined
by now and is described with the help of a model with a rather high accuracy
in the whole region of neutron binding energy by a ``smooth" function.
From our point of view, no alternative has been considered here.

However, the present-day fully model-free method of simultaneous
determination of $\rho$ and $k$ \cite{PEPAN-2005} gives another result.
Its practical application for more than 20 nuclei from the mass region
$(40 \leq A \leq 200)$ points to the existence in a nucleus of, at the least,
two excitation energies, in which abrupt and fundamentally important change
of its structure occurs. Approximation \cite{Prep196} of the experimental
data for $\rho$
by partial level densities of n-quasi-particle excitations shows that this
effect with maximum probability may be caused by the breaking of Cooper pairs
of nucleons in a heated nucleus with practically any mass.
Unfortunately, the lack of data \cite{Lone,pgaa} on the spectra of
gamma rays of radiative capture of thermal neutrons in $^{197}$Au has
prevented from using the method 
\cite{PEPAN-2005} to determine $\rho$ and $k$ in this nucleus. Both level
density and radiative strength functions in $^{198}$Au have been determined
\cite{Meth1}
using only the hypothesis of independence of the radiative strength functions
of primary and secondary gamma transitions of one and the same multipolarity
and energy on the excitation energy. The use of this assumption must
overestimate the $^{198}$Au level density determined experimentally in the
region of several MeV around 
$0.5B_n$ and underestimate the values of $k$ for the appropriate energies of
primary gamma transitions.
Estimation of the appropriate systematic error may be done on the basis of
comparison of the data for $\rho$ from \cite{PEPAN-2005,Meth1} for the nuclei
with a different parity of the number of neutrons and protons.
Relative smallness of the obtained error indicates that if we take the value
into account it will not lead to a fundamental distortion of conclusions of the
analysis of the available data.

Nevertheless, it is necessary to perform all possible analysis of the earlier
obtained experimental data \cite{Meth1} for this nucleus, in particular,
to estimate the degree of possible difference of radiative strength functions
for primary and secondary gamma transitions for various final levels of
$^{198}$Au, and also to estimate the degree of influence of other parameters
of this nucleus on experimental cascade intensities and the form of their
dependence on the intermediate level energy.

\section{Analysis}

It is very characteristic for the nuclei studied in accordance with the
methods \cite{PEPAN-2005,Meth1} that the change of sum $f(E1)+f(M1)$ at the
change of levels excited by them is of alternating-sign character:
a considerable increase of $k$ values in the region of ``stepped" structure
in the level density of relatively large energies excited by primary
transitions of levels is accompanied by some decrease of $k$ for low-lying 
cascade levels.
This effect in an odd-odd nucleus must lead to an overevaluation of calculated
cascade intensities at the increase of energy of their final level.

Two-step cascade intensities to the levels of $^{198}$Au with the energy
$E_f \leq 450$ keV required for comparison with the experiment \cite{Bon198}
have been calculated for the following variants of the level density:

a) the back-shifted Fermi gas model \cite{BSFG},

b) combination of the Ignatiuk model \cite{Ign} higher than 2 MeV with the
experimentally determined number of intermediate cascade levels at smaller
excitation energies and 

c) the experimental level density from method \cite{Meth1}.
In both variants of the method of simultaneous determination of $\rho$ and $k$
the level density of the positive and negative parity is varied independently.
However, at the same time the conservation of the average spacing for levels
corresponding to s-resonances and the summarized density of ``discrete" levels
is provided.

Radiative strength functions for E1-transitions are used from results
\cite{Meth1} and models 
\cite{KMF},\cite{Axel}. The model presentation for 
$f(M1)$ for the last two variants is restricted by the case 
$f(M1)=const$. The corresponding data is given in Fig. 1.

The summarized level density of both parities for spin window $1 \leq J \leq 3$
is presented in Fig. 2.

All the variants of the values $\rho$, $f(E1)$ and $f(M1)$ obtained in
accordance with \cite{Meth1} and presented in Fig. 1 and 2 practically
precisely reproduce the sum of cascade intensities to levels with the energy
less than 514 keV
\cite{Bon198} (Fig. 3). The comparison of experimental intensities of two-step
cascades to specific low-lying levels of $^{198}$Au for different variants
of level densities and radiative strength functions of the gamma transitions
is shown in Fig. 4a-c. The signs of random deviations of calculated intensities
in different sets of data from \cite{Meth1} anticorrelate with each other for
different final cascade levels.
The deviations of average values from the experimental intensities may be
partially related to the systematic errors of determination of the sums of
cascade intensities for the given final level.
Here, there is no reason to exclude a possible dependence of $I_{\gamma\gamma}$
values on $J^{\pi}_f$, and the influence of details of the structure of wave
functions of $E_f$ levels on the average value of 
$f(E1)$ and $f(M1)$ for secondary cascade transitions.

\section{Discussion of results}

In all the nuclei studied by now the final levels, the spin $J_f$ of which
differs from the neutron resonance spin $J_{\lambda}$ no more than for 2 and
has the maximum value at $|J_{\lambda}-J_f|=0$, are excited along with the
experimentally observed intensity.
For the compound state of $^{198}$Au excited at the thermal neutron capture
$J^{\pi}_{\lambda}=2^{+}$\cite{BNL}, therefore, the excess of cascade
intensities calculated according to the data 
\cite{Meth1} to the final levels 
$E_f=$347 and 406 keV with spins 
$J^{\pi}_f=2^{-}$ \cite{ENSDF} over the experimental values may be caused by
the presence of unresolved doublets and/or the influence of structure of the
enumerated levels on $f(E1)$ and $f(M1)$ of secondary transitions of the
cascades.
In any case this circumstance may not lead to changes of the obtained
conclusions on the process of cascade gamma decay due to the relative
smallness of excess of the calculated value in comparison with the value
$\sum_f I_{\gamma\gamma}$ observed in the experiment.

If we take into account such possible explanation then the calculation using
the data \cite{Meth1} gives a regular excess of intensity over the experiment
for cascades on the levels with the energy $261 <E_f < 482$ keV.
Due to the lack of other explanations it is possible to accept as the most
probable hypothesis that the relation of radiative strength functions of
secondary gamma transitions of cascades to the corresponding values for
primary ones has the same \cite{PEPAN-2005} form as in even-even nuclei and
in even-odd ones.
In other words, the general dynamics of the process of cascade gamma decay
of the neutron resonance is characterized by the regularities, which do not
depend on the parity of nucleon number in the odd-odd compound nucleus,
as well.

The presence of a local increase of radiative strength function in the region
of the ``stepped structure" in $\rho$
($\approx  1.5 <E_{ex} {^{<}_{\sim}} 3$ or for primary gamma transitions
($\approx 4 <E_{\gamma} {^{<}_{\sim}} 5-5.5$ MeV)
reflects, most likely 
\cite{PEPAN-2005}, a considerable increase of the influence of vibration
components of levels on its value in the region lower than the threshold
of appearance of four-quasi-particle excitations.
Radiative strength functions of primary gamma transitions decreased at the
breaking of subsequent Cooper pairs of nucleons to the levels with four-,
six-, etc. quasi-particle components.

In other words, new models of radiative strength functions must in an
explicit form take into account the co-existence and interaction of
excitations of quasi-particle and phonon types in the whole region under
consideration of excitation energy of a nucleus. In the level density models
this fact is explicitly taken into account, for example \cite{RIPL},
by introducing the vibration enhancement factor of level density of
quasi-particle type.
Therefore, most likely, no new types of excitation of nucleus
(of the ``pygmy-resonance" type) should be proposed and included in the $k$
models.

Almost the same value of the calculated cascade intensity in the energy region
$E_{ex} \approx 0.5B_n$ of their intermediate levels for all the tested
variants of radiative strength functions and level density demonstrates that
the conclusions made in \cite{LA172},\cite{Bud57},\cite{Pra198} by now on
the parameters of the process of cascade gamma decay must be in serious error,
since they do not take into account a strong correlation of values 
$\rho$ and $k$, which are included in $I_{\gamma\gamma}$.

In the framework of the existing notions it is impossible to reach the
conformity of the existing and possible models of $\rho$ and $k$ with the
experiment by any parameter variation, if only they do not take into account
quite realistically the influence of breaking of the nucleon Cooper pairs on
these parameters of the process of cascade gamma decay of the high-excited
level.

\section{Conclusion}

Currently, there are no obstacles in obtaining the experimental data necessary
for a rather detailed theoretical description of the properties of nucleus
lower than $\approx B_n$.
By analogy with the experience of study of two-step cascades at the thermal
neutron capture one may assume that reliable data on $\rho$ and $k$ in other
experiments may be obtained only at the study of two-step nuclear reactions
in coincidences by high-resolution spectrometers.

Erroneous conclusions during the analysis of an experiment of such type may
occur only if out-dated model notions on the level density or the probability
of emission of nuclear reaction products are used.

The potential models of the level density of a nucleus and radiative strength
functions of gamma transitions exciting them must in an explicit form take
into account the co-existence and interaction of excitations of quasi-particle
and phonon type at least lower than the neutron binding energy.
Practical necessity in their development became apparent \cite{Maslov} at
the evaluation of contemporary data on cross sections of the interaction of
neutrons with fissionable nuclei.

\begin{figure}

\vspace{2cm}
\begin{center}
\leavevmode
\epsfxsize=16.5cm
\epsfbox{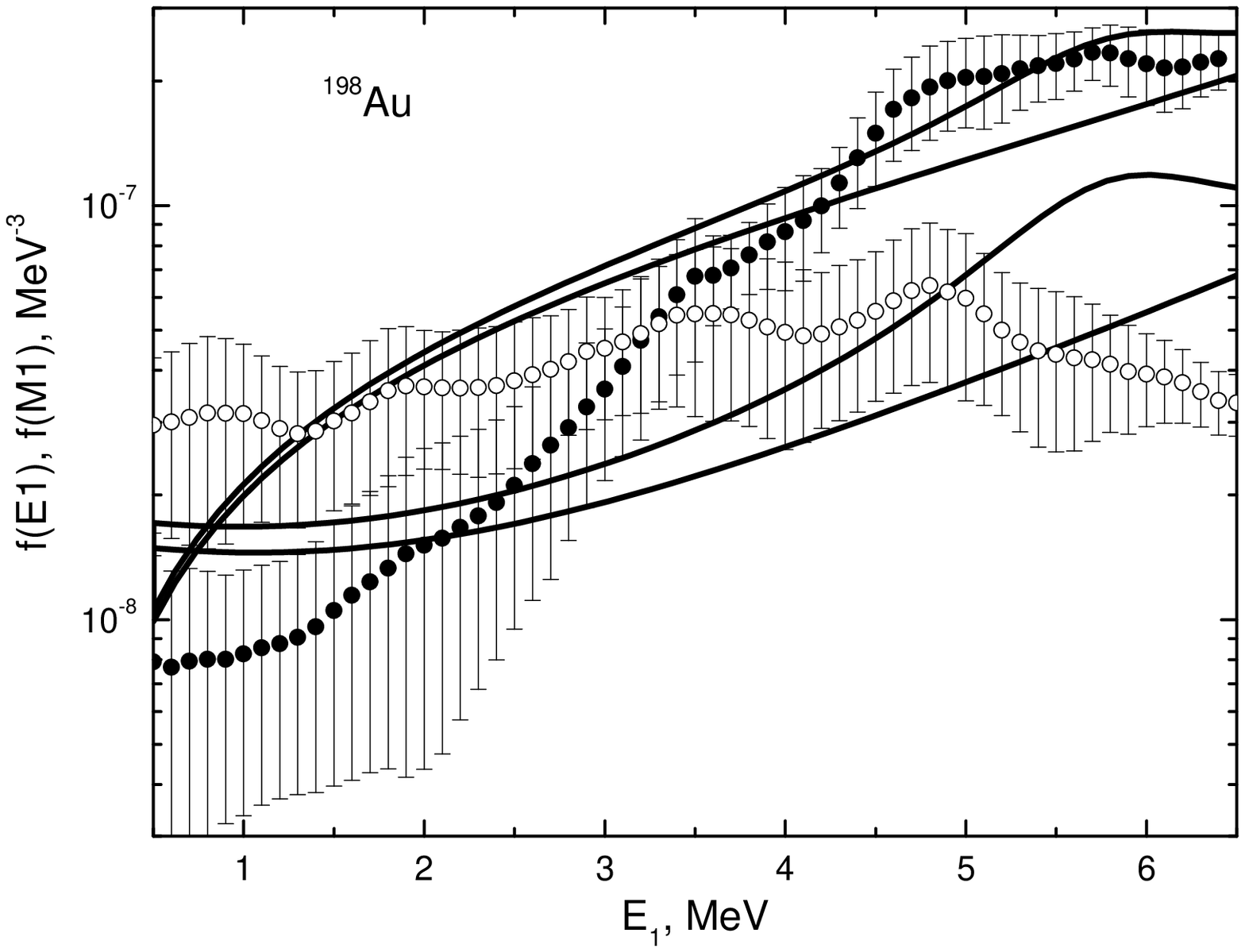}
\end{center}
\vspace{-8cm}
{\bf Fig.~1.} Solid thick lines - $f(E1)$ from models \cite{KMF,Axel}
and their sums
with ``pygmy-resonance" for its parameters from \cite{Pra198}.
Open point with errors -- region for set of random functions of 
$f(M1)$, solid points -- $f(E1)$, reproducing 
$I_{\gamma\gamma}$ (Fig. 3) with practically the same values $\chi^2/f <<1$.
\end{figure}

\begin{figure}

\vspace{1cm}
\begin{center}
\leavevmode
\epsfxsize=16.5cm
\epsfbox{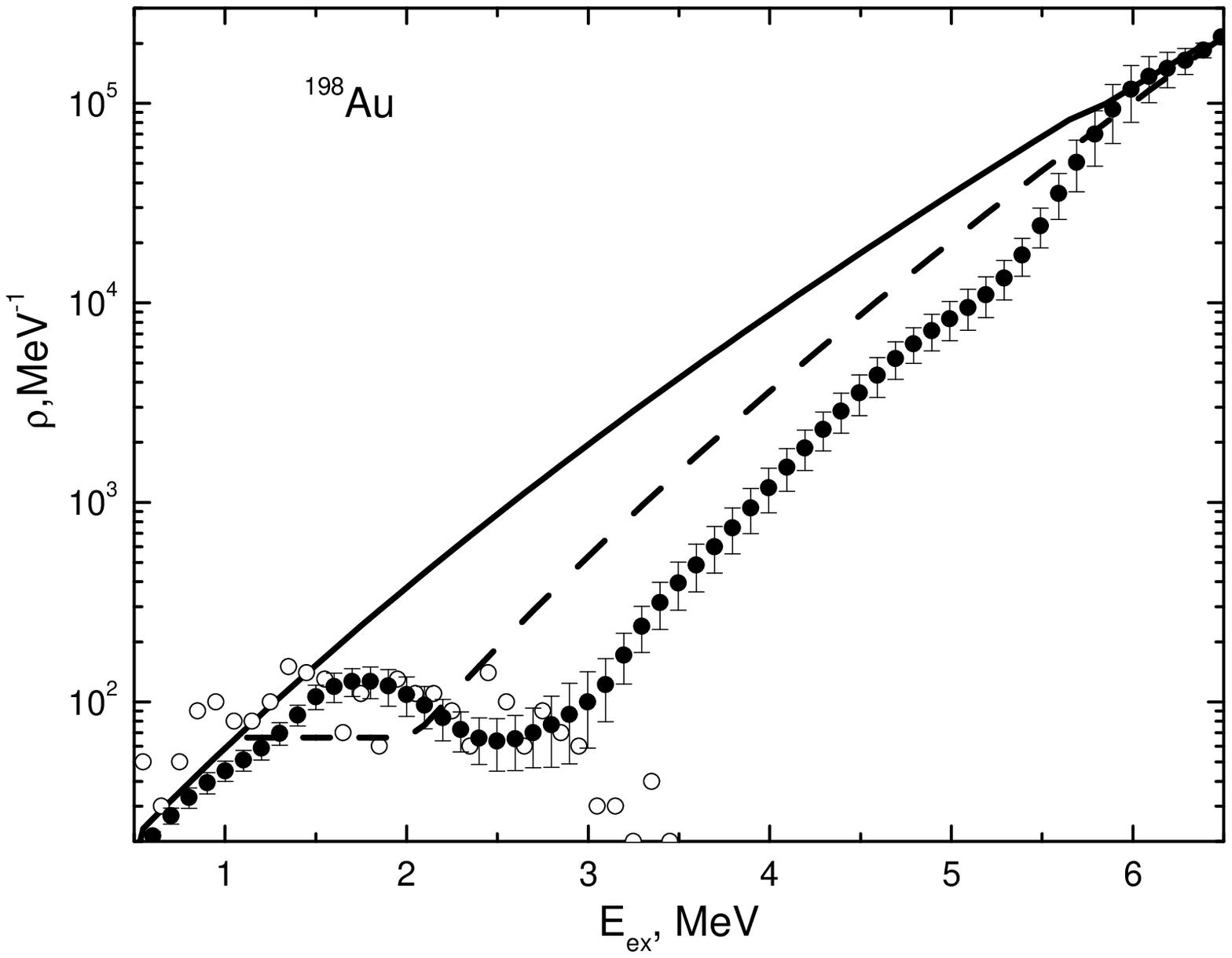}
\end{center}
\vspace{-6cm}

{\bf Fig.~2.} The same as in Fig. 1 for the level density
(solid points with errors).
Solid line -- model values \cite{BSFG}, dotted line -- \cite{Ign}
respectively.
Open points -- the density of intermediate cascade levels \cite{Meth1}.
\end{figure}
\newpage
\begin{figure}

\begin{center}
\leavevmode
\epsfxsize=16.5cm
\epsfbox{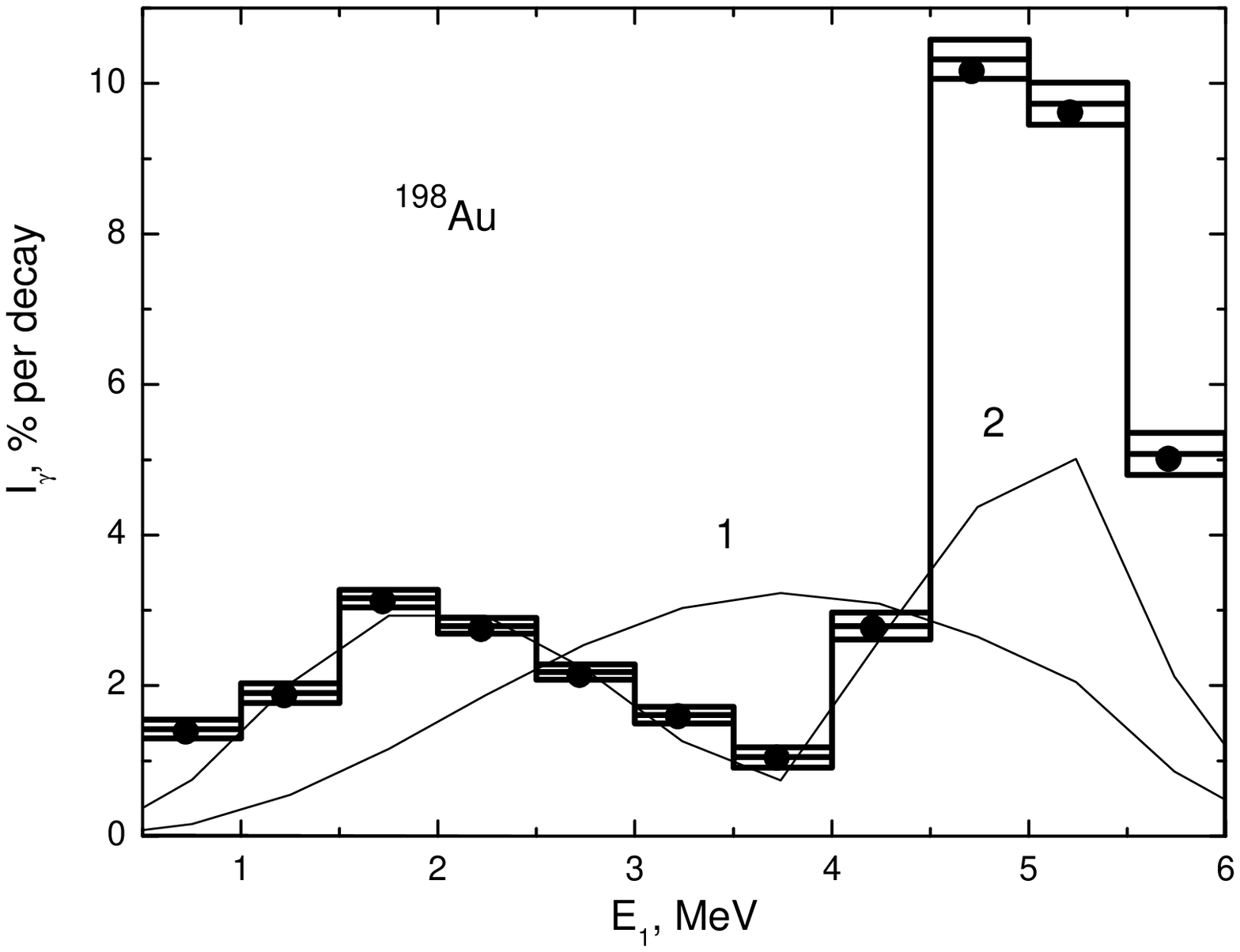}
\end{center}
\vspace{-6cm}
{\bf Fig.~3.} Hystogram -- summarized experimental intensity of two-step
cascades in the intervals of 0.5 MeV
in the function of energy of the primary gamma transition with statistical
errors only \cite{Bon198}.
Line 1 -- calculation with level density from \cite{BSFG}, line 2 -- \cite{Ign}.
Points -- the typical approximation for the data from \cite{Meth1},
the examples of which are given in Fig. 1 and 2.
\end{figure}
\begin{figure}

\begin{center}
\leavevmode
\epsfxsize=16.5cm
\epsfbox{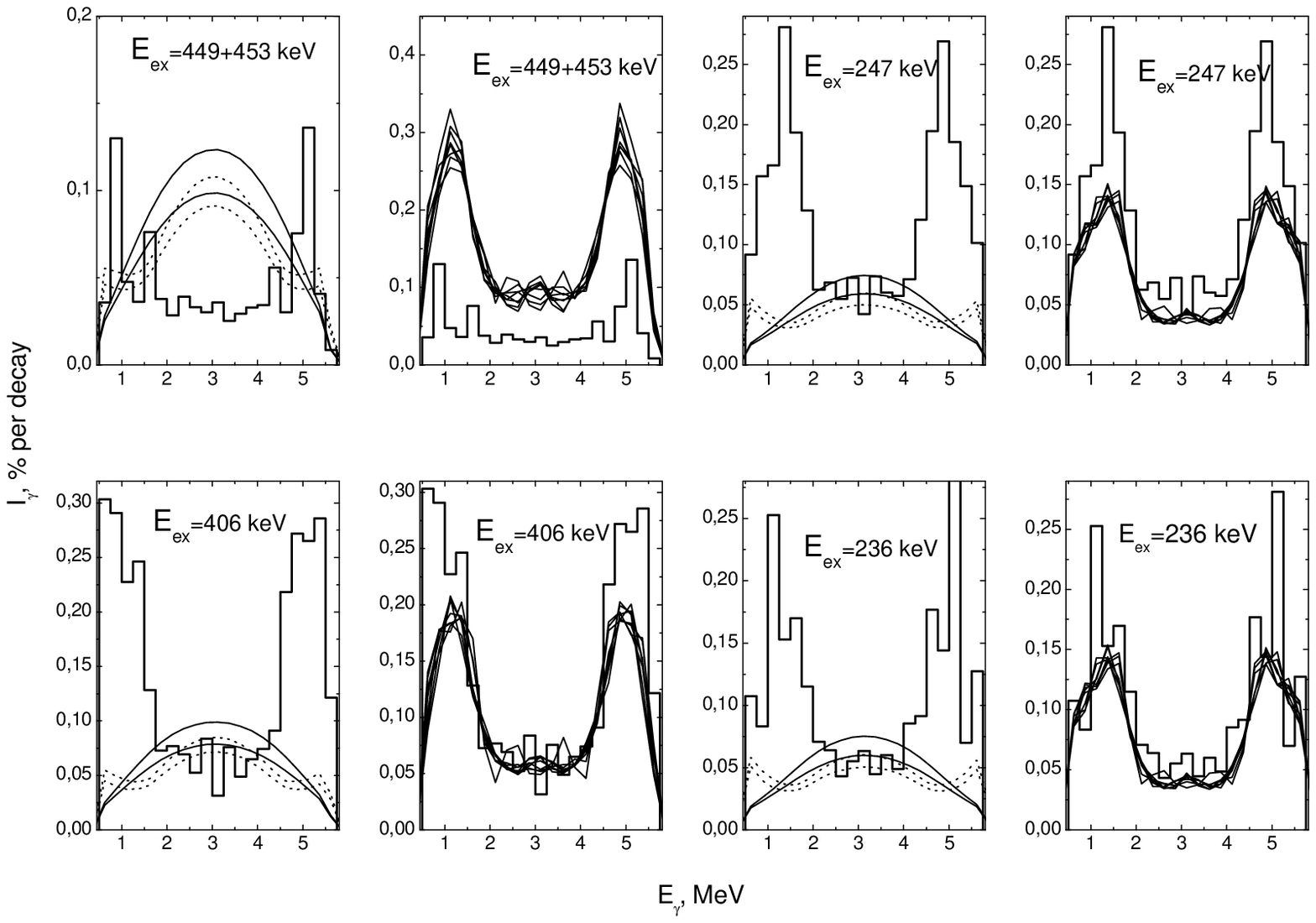}
\end{center}
\vspace{-8cm}

{\bf Fig.~4a.} Hystogram -- experimental intensity of two-step cascades for
the levels $E_{ex}$ (summed over the intervals of 250 keV),
lines -- different variants of the calculation.
The first and third columns: combinations of models \cite{BSFG,KMF,Axel} --
thin lines,
\cite{Ign,KMF,Axel} -- dotted  line. The second and fourth columns - 
the intensity is calculated for random sets of $\rho$ and $k$ from
\cite{Meth1}. 
\end{figure}
\begin{figure}

\vspace{-3cm}
\begin{center}
\leavevmode
\epsfxsize=16.5cm
\epsfbox{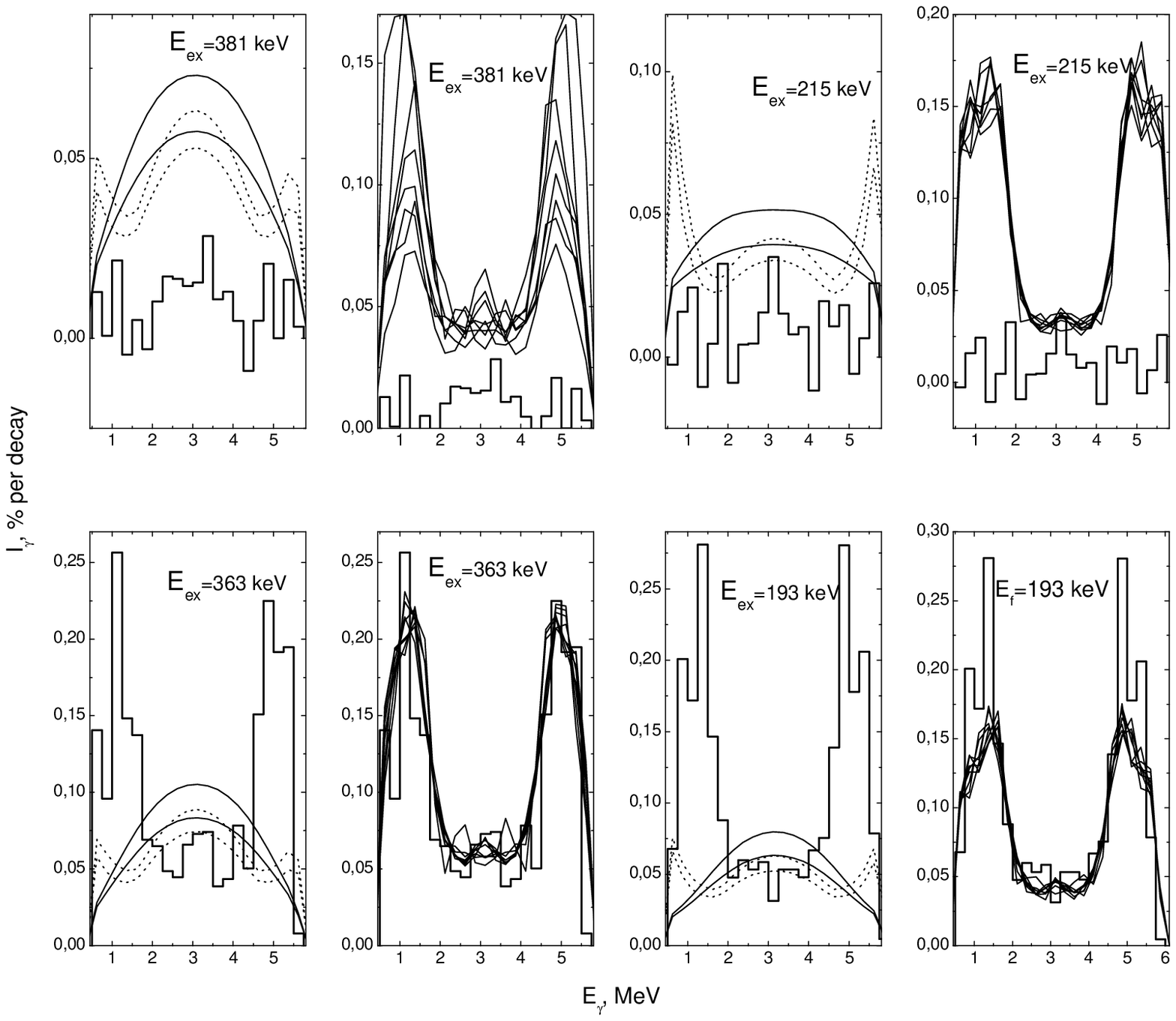}
\end{center}
\vspace{-8cm}

{\bf Fig.~4b.}  The same as in Fig.4a for other final levels.
\end{figure}
\newpage
\begin{figure}

\vspace{-1cm}
\begin{center}
\leavevmode
\epsfxsize=16.5cm
\epsfbox{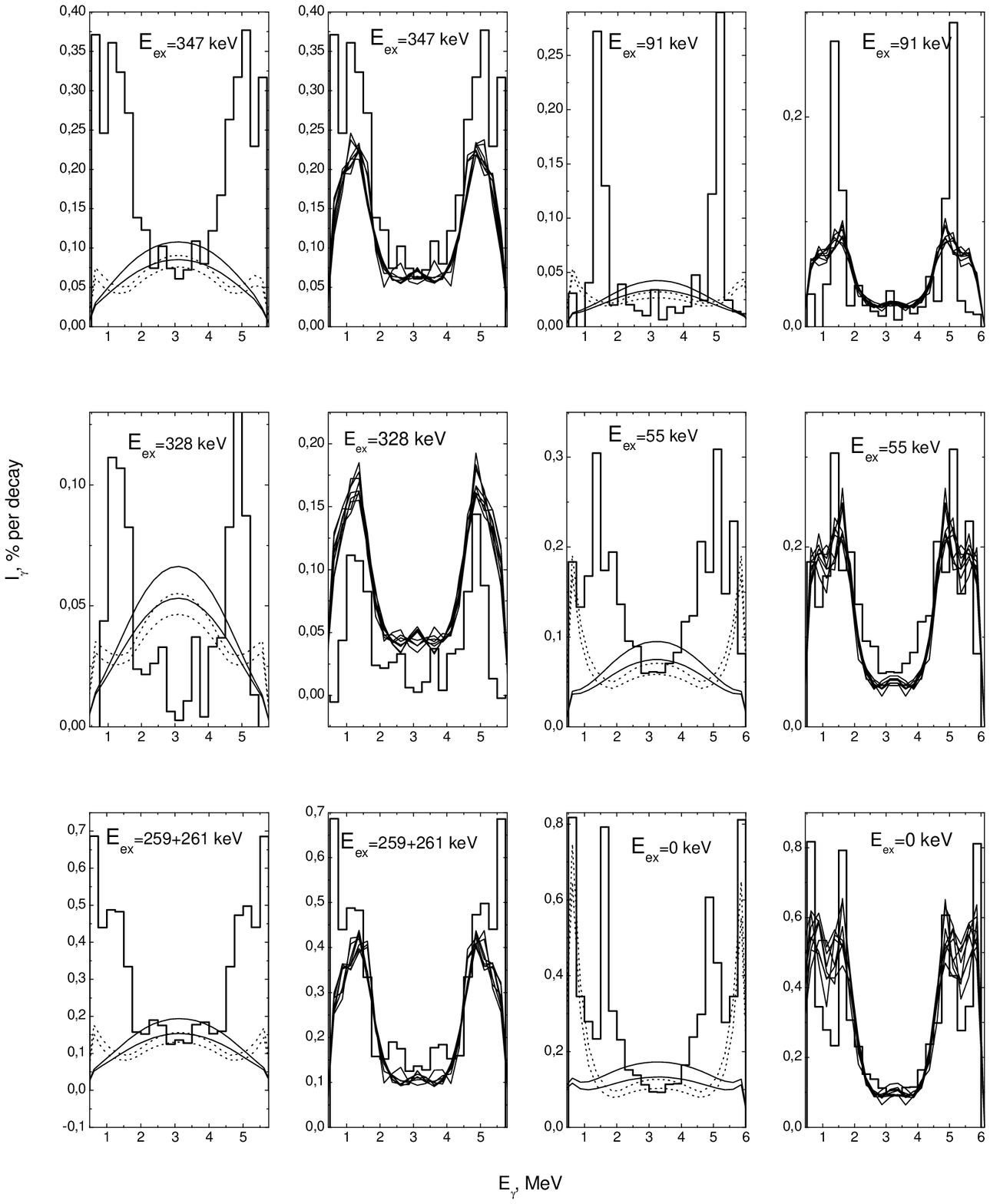}

\end{center}

\vspace{-5cm}
{\bf Fig.~4c.} The same as in Fig.4a for other final levels.
\end{figure}

\end{document}